\documentclass[a4paper,10pt]{article}
\usepackage{natbib}
\usepackage{epsfig}
\usepackage{amsmath}
\usepackage{multicol}

\newcommand{\avk}{\langle k \rangle}
\newcommand{\fluck}{\langle k^2 \rangle}
\newcounter{Sectioncounter}
\newcounter{SubSectioncounter}

\newcommand{\Section}[1]
{ {
\vspace{10pt}
\addtocounter{Sectioncounter}{1}
\setcounter{SubSectioncounter}{0}
\begin{center}
{\bf \arabic{Sectioncounter}. #1}
\end{center}
} }
\newcommand{\Subsection}[1]
{ {
\vspace{5pt}
\addtocounter{SubSectioncounter}{1}
\begin{center}
{\sc \arabic{Sectioncounter}.\arabic{SubSectioncounter}. \lowercase{#1}}
\end{center}
} }
\newcommand{\Appendix}[1]
{ {
\vspace{20pt}
\begin{center}
{\bf Appendix}
\end{center}
} }

\makeatletter

\newenvironment{figurehere}
  {\def\@captype{figure}}
  {}
\makeatother

\renewcommand{\baselinestretch}{1}

\begin{document}

\begin{center}
  {\large \bf Dynamical patterns of epidemic  outbreaks in complex 
heterogeneous networks}\\
  
  \vspace*{0.5cm} 
  
  {\small \sc Marc Barth{\'e}lemy$^1$, Alain Barrat$^2$, Romualdo
    Pastor-Satorras$^3$, and Alessandro Vespignani$^{2,4}$}

  \vspace*{0.5cm}

  {\em $^1$CEA-Centre d'Etudes de Bruy{\`e}res-le-Ch{\^a}tel, D{\'e}partement
    de Physique Th{\'e}orique et Appliqu{\'e}e BP12,\\ 91680
    Bruy{\`e}res-Le-Ch{\^a}tel, France\\ $^2$Laboratoire de Physique
    Th{\'e}orique (UMR du CNRS 8627), B\^atiment 210\\ Universit{\'e} de
    Paris-Sud 91405 Orsay, France\\ $^3$Departament de F\'isica i
    Enginyeria Nuclear, Universitat Polit{\`e}cnica de Catalunya\\
    Campus Nord, 08034 Barcelona, Spain\\ $^4$School of Informatics and
    Biocomplexity center, Indiana University, Bloomington IN 47408 USA.
    
    \vspace*{0.5cm} \rm (\today)
}

\end{center}

\begin{quotation}
  We present a thorough inspection of the dynamical behavior of epidemic
  phenomena in populations with complex and heterogeneous connectivity
  patterns. We show that the growth of the epidemic prevalence is
  virtually instantaneous in all networks characterized by diverging degree
  fluctuations, independently of the structure of the connectivity
  correlation functions characterizing the population network.  By means of
  analytical and numerical results, we show that the outbreak time evolution
  follows a precise hierarchical dynamics.  Once reached the most highly
  connected hubs, the infection pervades the network in a progressive cascade
  across smaller degree classes. Finally, we show the influence of the initial
  conditions and the relevance of statistical results in single case studies
  concerning heterogeneous networks.  The emerging theoretical framework
  appears of general interest in view of the recently observed abundance of
  natural networks with complex topological features and might provide useful
  insights for the development of adaptive strategies aimed at epidemic
  containment.
\end{quotation}

\vspace*{0.25cm}

\begin{multicols}{2}

\Section{Introduction}

Accurate mathematical models of epidemic spreading are the basic
conceptual tools in understanding the impact 
of diseases and the development of effective strategies for their 
control and containment~\citep{bailey1975,anderson92,diekbook,daileybook}. 
In order to increase the
relevance and utility of these strategies, it is crucial 
that models contain the key features that appropriately characterize
the system of interest. The age and social structure of the
population, the contact network among individuals, and the metapopulation
characteristics such as the geographical patch structure, are all
factors that might acquire a particular relevance in a reliable
epidemic model (for a recent review on the subject 
see ~\citep{Ferguson:2003}).  
Among these features, the connectivity pattern of the network of 
contacts among individuals along which
the disease can be transmitted has been acknowledged since a long time
as a relevant factor in determining the properties of epidemic spreading
phenomena 
(see \cite{yorkesuperspreader,het84,may88,anderson92,Eubank:2004} 
and references
therein). In these studies, it was observed that the heterogeneity 
of the population network  in  which the disease
spreads may  have noticeable effects in the evolution of the epidemic
as well as in the effect of immunization strategies. 

The study of the impact of network structure in the epidemic modeling
has been recently revamped by the large number of results on large
networked systems pointing out the ubiquitous presence of
heterogeneities and complex topological features on a wide range of
scales \citep{barabasi02,mendesbook,romuvespibook}. A striking example of
this new framework is provided by scale-free networks, which are
characterized by virtually infinite fluctuations in the number of 
connections $k$ (the degree) that any given vertex in the network may
have.  This feature finds its signature in a heavy-tailed degree 
distribution (defined as the probability that
any vertex is connected to $k$ other vertices), often approximated by
a  power-law behavior of the form $P(k)\sim k^{-\gamma}$, 
with $2<\gamma\leq 3$. This implies an unexpected statistical
abundance of vertices with very large degrees; 
i.e. the so called ``hubs'' or ``superspreaders''
\citep{yorkesuperspreader}. In scale-free networks the abundance 
of those vertices reaches a level that guarantees the proliferation of a
large number of infected individuals whatever the rate of infection
characterizing the epidemic, eventually leading to the absence of
any epidemic threshold below which the infection cannot initiate a
major outbreak~\citep{pv01a,pv01b,may01,moreno02,newman02}.
This peculiar theoretical scenario, however, turns out to be of 
practical importance since both the sexual contact pattern
\citep{colgate89,amaral01,schnee04} and several technological networks  
\citep{pv01a,lloyd01,romuvespibook} appear to have scale-free
features. This implies that both sexually transmitted diseases and
computer viruses may fit in this scenario, raising new questions
and scrutiny on several epidemic models and strategies 
aimed at optimizing the deployment of immunization 
resources~\citep{aidsbar,havlin03}.

In this paper we provide a thorough presentation of results concerning
the analysis of the time evolution of epidemic outbreaks in complex
networks with highly heterogeneous connectivity patterns. We consider
the time behavior of epidemic outbreaks in the general class of models
without and with internal recovery and find that the growth of
infected individuals is governed by a time scale $\tau$ proportional
to the ratio between the first and second moment of the network's
degree distribution~\citep{barthelemy04}, $\tau\sim \avk/\fluck$.
This implies that the larger the degree fluctuations (governed by
$\fluck$), the faster the epidemic propagation will be.  In
particular, a virtually instantaneous rise of the prevalence is
obtained in scale-free networks where $\fluck\to\infty$ for infinite
network sizes. This result is shown to be valid also in networks with
non trivial connectivity correlation functions as often encountered
in real systems analysis. Furthermore, we study the detailed
propagation in time of the infection through the different degree
classes in the population. We find a striking hierarchical dynamics in
which the infection propagates via a cascade that progresses from
higher to lower degree classes. This infection hierarchy might be used
to develop dynamical ad-hoc strategies for network
protection. Finally, we study the influence of initial conditions on
the epidemic development and the relevance of statistical results in
the case of single case studies. All the analytical discussion is
supplemented with careful numerical simulations at the discrete
individual level.

%%%%%%%%%%%%%%%%%%%%%%%%%%%%% State of the art in dynamics
\Section{Basic theory of epidemic dynamics}

\Subsection{Reproductive number and time evolution}

At first instance epidemiological studies deal with the properties of
epidemics in the equilibrium or long time steady state such as a
non-zero prevalence state associated to the presence of an endemic
phase, the presence or absence of a global outbreak, or the
non-seasonal cycles that are observed in many infections
\citep{anderson92}. As well, the dynamical evolution of 
epidemics outbreaks and the effects of the introduction of a seed of
infection into a large population of susceptible individuals are of
great concern. A basic parameter in epidemiology is the basic
reproductive number $R_0$, which counts the number of secondary
infected cases generated by one primary infected individual.  Under
the assumption of the homogeneous mixing of the population if an
infected individual is in contact with $\avk$ other individual, the
basic reproductive number is defined as
\begin{equation}
  R_0 = \frac{\lambda \avk}{\mu},
\end{equation}
where $\lambda$ is the spreading rate, defined as the probability rate that
a susceptible individual in contact with an infected individual will
contract the disease, and $\mu$ is the recovery rate of infected
individuals, either to the susceptible or the recovered states.  
It is easy to understand that any epidemic will
spread across a non zero fraction of the population only for $R_0>1$.
In this case the epidemic is able to generate a number of infected
individuals larger than those which are recovered,  leading to an
increase of the infected individuals $i(t)$ at time $t$ 
following the exponential form 
\begin{equation}
i(t)\simeq i_0e^{t / \tau_d}.
\label{eq:ave_i}
\end{equation}
Here $i_0$ is the initial density of infected individuals and $\tau_d$ is
the typical outbreak time, that in general reads as \citep{anderson92}
\begin{equation}
  \tau_d^{-1}=\mu  (R_0-1).
\end{equation}
The previous considerations lead to the definition of a crucial
epidemiological concept, namely the epidemic threshold. Indeed, if the
spreading rate is not large enough to allow a reproductive number
larger than one ($\lambda>\mu/\avk$), the epidemic outbreak will not
affect a finite portion of the population and dies out in a finite
time.  In epidemiological studies for which the two important
assumptions of ``homogeneous mixing'' and constant infectiousness
(constant $\lambda$) are made, the spreading pattern of the epidemics
is therefore controlled by the generation time scale $1/\lambda$ and
$R_0$ and there are roughly three different stages in an epidemics
\citep{Ferguson:2003}.  More precisely, when infectious individuals
are introduced in a network, one observes a first noisy phase followed
in general by an exponential outbreak of the epidemics.  Depending on
the long term behavior of individuals against the disease we will
observe at large times a different behavior described by the specific
epidemic model used (see Fig.~(\ref{fig:fluctuat})).

\begin{figurehere}
\vspace*{1cm}
\centerline{
\epsfysize=0.5\columnwidth{\epsfbox{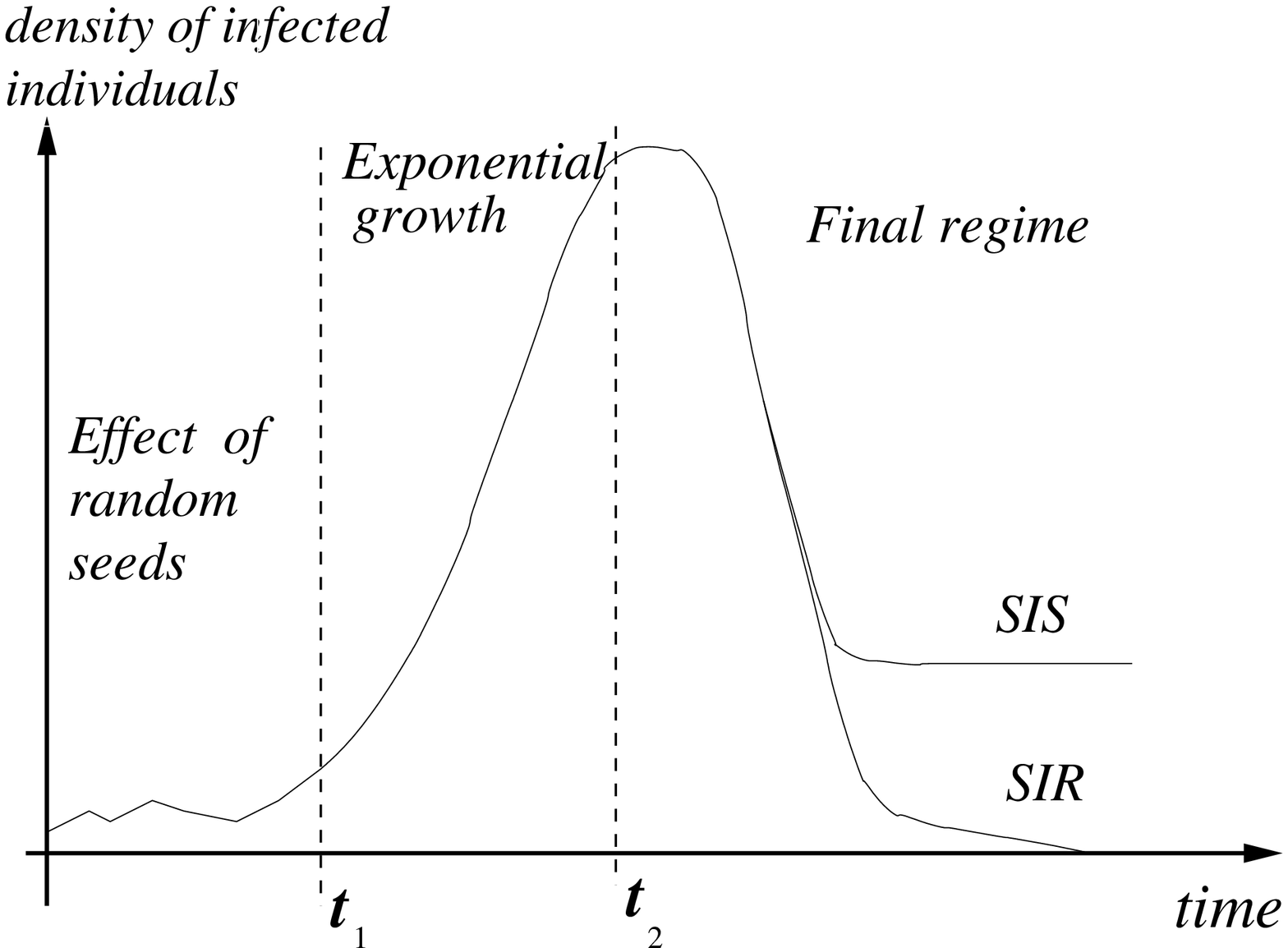}}
}
\vspace*{0.1cm}
\caption{ \small Typical profile of density of infected versus time on a given
  realization of the network. In the first regime $t<t_1$, the
  outbreak did not start and there are fluctuations. In the second
  regime, $t_1<t<t_2$ there is an exponential growth characterized by
  $R_0$. In the final regime ($t>t_2$), the density of infected either
  converges to a constant for the SIS model or to zero for the SIR
  model.  } \vspace*{1cm}
\label{fig:fluctuat}
\end{figurehere}

The above considerations and parameters are at the core of several
epidemic models based on the compartmentalization of the population.
In other words, each individual of the population can only exist in a
certain number of discrete states such as susceptible, infected or
permanently recovered. The latter state is equivalent to the removal
of the individual from the population since it is supposed that it
cannot get the infection anymore. The total population $N$ is assumed
to be constant and if $S(t)$, $I(t)$ and $R(t)$ are the number of
susceptible, infected and removed individuals at time $t$,
respectively, then $N=S(t)+I(t)+R(t)$.  The simplest epidemiological
model one can consider is the susceptible-infected-susceptible (SIS)
model.  The SIS model is mainly used as a paradigmatic model for the
study of infectious diseases leading to an endemic state with a
stationary and constant value for the prevalence of infected
individuals, i.e. the degree to which the infection is widespread in
the population. In the SIS model, individuals can only exist in two
discrete states, namely, susceptible and infected.  The disease
transmission is described in an effective way.  The probability that a
susceptible vertex acquires the infection from any given neighbor in
an infinitesimal time interval $d t$ is $\lambda d t$, where $\lambda$
defines the virus \index{spreading rate}\textit{spreading rate}.  At
the same time, infected vertices are cured and become again
susceptible with probability $\mu d t$.  Individuals thus run
stochastically through the cycle susceptible $\to$ infected $\to$
susceptible, hence the name of the model.  The SIS model does not take
into account the possibility of individuals removal due to death or
acquired immunization, which would lead to the so-called
susceptible-infected-removed (SIR) model \citep{anderson92,murray}.
The SIR model, in fact, assumes that infected individuals disappear
permanently from the network with rate $\mu$ .  In models such as the
SIS, the number of infected individuals increases up to a stationary
constant value which is non zero if $R_0>1$.  On the contrary, in
models such as the SIR, the number of infected individuals tends
toward zero since all infected will sooner or later become removed
from the population.  Also in this case, however, a finite fraction of
the population is affected by the epidemic outbreak only if $R_0>1$.
It should be noted that it is also possible to induct a steady state
in the SIR model, by introducing new susceptible individuals at a
constant rate.  This new parameter constitutes a new time scale that
gives rise to oscillations in the endemic phase \citep{may84}.

\Subsection{Complex heterogeneous networks}

The general picture presented in the previous section is obtained 
in the framework of the
homogeneous mixing hypothesis. This hypothesis assumes that the
network of contacts among individuals has very small degree 
fluctuations. In other
words, the degree $k$ fluctuates very little and we can assume
$k\simeq \avk$, where the brackets $\langle\cdot\rangle$ denote the
average over the degree distribution.

However, networks can be very heterogeneous. Social heterogeneity and
the existence of ``super-spreaders'' have been known for long time in
the epidemics literature~\citep{het84}. The signature of this large
heterogeneity can be measured in the degree distribution $P(k)$ which
in this case decays very slowly. Indeed, in a homogeneous network such
as a random graph \citep{erdos59}, $P(k)$ decays faster than
exponentially, while for scale-free networks \citep{barabasi02} it
decays as a power law for large $k$
\begin{equation}
P(k)\sim k^{-\gamma}.
\end{equation}
Examples of such networks relevant to epidemics studies include the
Internet \citep{romuvespibook}, the network of airline
connections~\citep{mossa2,barratairport}, or the web of sexual
contacts~\citep{colgate89,amaral01,schnee04}.  In these networks, the average 
degree $\avk$ is no longer the relevant variable and one expects the
fluctuations, described by $\fluck$, to play an important role.

The question of the effect of the heterogeneity on epidemic behavior 
has been addressed at various levels~\citep{het84} and analyzed in 
details in the last years for scale-free networks~\citep{pv01a,pv01b}. 
These studies were concerned with the
stationary limit and the existence of an endemic phase. A key result
is the expression of the basic reproductive number which in this case
takes the form
\begin{equation}
R_0\propto\frac{\fluck}{\avk} \ .
\end{equation}
The important fact here is that $R_0$ is proportional to the second
moment of the degree, which diverges for increasing network sizes.
This has some important epidemiological consequences. Indeed, whatever
the spreading rate $\lambda$ the basic reproductive rate is always
larger than one, thus leading to the lack on any epidemic threshold. 
In other words, in heterogeneous networks, whatever the 
infection rate, the epidemics has a finite probability to generate a 
major outbreak. 

These results, while extremely important, consider only one face of
networks complexity. In general, however, real networks have other
complex features such as strong degree correlation among connected vertices.
This feature is mathematically characterized through the 
conditional probability $P(k'|k)$
that a vertex of degree $k$ is connected to a vertex of degree $k'$.
Networks which are completely defined by the
degree distribution $P(k)$ and the conditional probability $P(k'|k)$
are called \textit{Markovian networks} \citep{marian1}
and must fulfill the following degree detailed balance
condition
\begin{equation}
  k P(k'|k) P(k) = k' P(k|k') P(k').
  \label{eq:1}
\end{equation}
This expression is a mathematical statement of the obvious observation that, in
any real network, all edges must point from one vertex to another. 

The full knowledge of the function $P(k'|k)$, which measures
the correlations in the network, would often be
difficult to interpret. Therefore, the {\em average nearest neighbor degree}
$k_{nn}$, and the behavior of this quantity as a function of the
degree, $k_{nn}(k)=\sum_{k'} k'P(k'|k)$, have been proposed to measure
these correlations~\citep{romuvespibook}. In the absence of
degree correlations, $P(k'|k)$ does not depend on $k$ and neither does
the average nearest neighbors' degree; i.e. $k_{nn}(k)= {\rm const}.$
\citep{romuvespibook}. In the presence of correlations, the
behavior of $k_{nn}(k)$ identifies two general classes of networks. If
$k_{nn}(k)$ is an increasing function of $k$, vertices with high
degree have a larger probability to be connected with large degree
vertices. This property is referred in physics and social sciences as
{\em assortative mixing} \citep{assortative}. On the contrary, a
decreasing behavior of $k_{nn}(k)$ defines {\em disassortative
mixing}, in the sense that high degree vertices have a majority of
neighbors with low degree, while the opposite holds for low degree
vertices. These possibilities are summarized in Figure \ref{fig:knn}.
Among real networks, many social networks display assortative mixing, while
technological networks show typically disassortative properties.
In summary, the average nearest neighbor degree carry an information
about two-point correlations which is easy to interpret and avoids
the fine details of the full distribution $P(k'|k)$.

\begin{figurehere}
\vspace*{1cm}
\centerline{
\epsfysize=0.5\columnwidth{\epsfbox{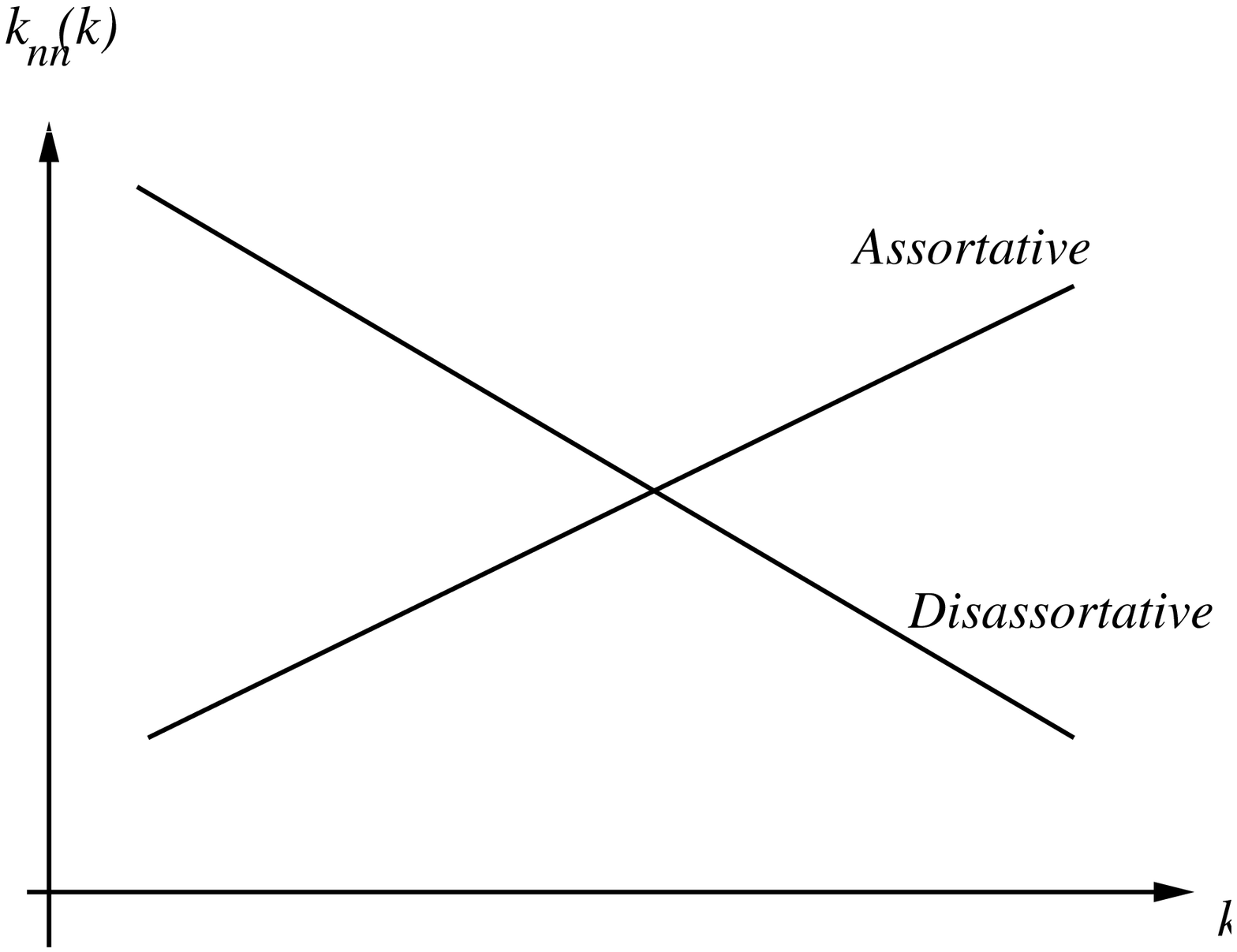}}
}
\vspace*{0.1cm}
\caption{ \small Typical behavior of the average
nearest neighbor degree $k_{nn}(k)$ for assortative mixing and
disassortative mixing.} \vspace*{1cm}
\label{fig:knn}
\end{figurehere}

The propagation of epidemics on a network may {\em a priori} be
modified by the presence of correlations. The quantity $k_{nn}$
seems in this context very relevant, since it measures the
number of individuals that may be reached by the infection in few steps.
In general, however, it has been shown in \citep{boguna:corre} that, in any
scale-free network with diverging second moment $\fluck$, the epidemic
threshold of the model vanishes, regardless of the presence of
correlations. In finite networks, the role of correlations appears
through the fact that the epidemic threshold is bounded from above by
a quantity which depends on $k_{nn}$~\citep{boguna:corre}.

While these results contribute to a general understanding of
epidemic spreading in complex networks, they do not provide a large
amount of information 
about the time patterns of the dynamical process and how effectively 
the spreading occurs.
Uncovering these patterns and their relation to the
heterogeneity of the networks is crucial for assessing control
strategies. In the following sections we will approach this problem
within a general formalism, valid for most kind of epidemic processes
in heterogeneous networks.

%%%%%%%%%%%%%%%%%%%%%%%%% short time behavior: typical outbreak time
\Section{The early stages of the  epidemic outbreak}

In order to understand how the topology of the network of contacts
affects the dynamical process of epidemics spreading, we will perform
a study on different kinds of networks.  Most complex networks can be
classified according to the decay of their degree distribution
\citep{amaral}. We will first consider the case where the degree
distribution decays faster than any power-law, typically as an
exponential (or faster), such as for the random (Poisson) graph
\citep{erdos59}. For this sort of networks, the degree fluctuations
are very small and we can approximate the degree of every vertex as a
constant, $k\approx \avk$. This approximation, used in most previous
studies \citep{anderson92,murray}, corresponds to the homogeneous
mixing case in which all individuals have the same environment, and
the same number of acquaintances that can be infected. We will then
consider the case of heterogeneous networks, exemplified in the case
of the scale-free networks, and show how the large heterogeneity
induces changes in the epidemic process.

Both the SIS and SIR models introduce a time scale $\tau_r=1/\mu$
governing the self-recovery of individuals. Since we are especially 
interested in the onset dynamics of outbreaks two scenarios 
are in order. If $\tau_r$ is smaller than the spreading time scale, 
then the process is dominated by the natural recovery of infected to 
susceptible or removed individuals. This situation is however 
less interesting  since it corresponds to a dynamic process 
governed by the decay into a healthy state and the interaction 
with neighbors plays a minor role. Epidemiological concern is
therefore  in the regime $\tau_r\gg
1/\lambda$; i.e. a spreading time scale much smaller than 
the recovery time scale. In this case, as a first approximation, we
can neglect the individual recovery that will occur at a much later
stage and focus on the early dynamics of the epidemic outbreak. 
This case corresponds to the simplified susceptible-infected (SI) 
model, for which infected nodes remain always infective and spread 
the infection to susceptible neighbors with rate $\lambda$;
the seed of infectives placed at time $t=0$ will thus infect the rest
of the network, the dynamical process being controlled by the topology
of the network. The SI model thus appears as the simplest framework 
for assessing the
effect of network's topology on the spreading dynamics.  This argument
is the reason why we will present our calculations for the simple SI
model. For the sake of completeness, however, we will readily show
in the next sections that the results obtained for the SI model can be
considered as fairly general and can be easily extended to SIS
and SIR models for heterogeneous networks.

%%%% random graph
\Subsection{Homogeneous networks}

A first analytical description of the SI model can be undertaken
within the homogeneous mixing hypothesis~\citep{anderson92,murray},
consisting in a mean-field description of the system in which all
vertices are considered as being equivalent. In this case the system
is completely defined by the number of infected individual $I(t)$, and
the reaction rate equation for the density of infected individuals
$i(t)=I(t)/N$ (where $N$ is the total size of the population) reads as
\begin{equation}
  \frac{d i(t)}{dt}=\lambda \avk i(t)\left[1-i(t)\right].
  \label{SI-MF}
\end{equation}
The above equation states that the growth rate of infected individuals
is proportional to the spreading rate $\lambda$, the density of
susceptible vertices that may become infected, $s(t)=1-i(t)$, and the
number of infected individuals in contact with any susceptible vertex.
The homogeneous mixing hypothesis considers that this last term is
simply the product of the number of neighbors $\avk$ and the average
density $i(t)$. Obviously, this approximation neglects correlations
among individuals and considers that all vertices have the same number
of neighbors $\avk$, i.e. it assumes a perfectly homogeneous network.
The solution of equation (\ref{SI-MF}) reads
\begin{equation}
i(t)= \frac{i_0 \exp(t/\tau_H)}{1 + i_0 [\exp(t/\tau_H) -1]} \ ,
\label{i_SI_H}
\end{equation}
where $i_0$ is the initial density of infected individuals and
$\tau_H= (\lambda \avk)^{-1}$ is the time scale of the infection
growth.  At small times, when the density of infected vertices is very
small, the leading behavior is given by $i(t)\simeq i_0
e^{t/\tau_{H}}$, i.e.  by an exponential growth. The value of $\tau_H$
corresponds to the intuitive fact that small spreading rates decrease
the growth velocity, while a larger number of neighbors ensures a
faster invasion of the network.

%%% scale free network
\Subsection{Heterogeneous networks}

The above calculations are valid for networks in which the degree
fluctuations are very small, i.e. $k \approx \langle k \rangle$ for
all vertices in the network.  We face however a different situation in
networks with a heterogeneous connectivity pattern. In this case the
degree $k$ of vertices is highly fluctuating and the average degree is
not anymore a meaningful characterization of the network properties.
In order to take this into account, it is possible to write down the
reaction rate equations for the densities of infected vertices of
degree $k$, $i_k(t)=I_k(t)/N_k$, where $N_k$ (resp. $I_k(t)$) is the
number of vertices (resp. infected vertices) within each degree class
$k$~\citep{pv01a}. In the case of the SI model the evolution equations
read \citep{marianproc}
\begin{equation}
  \frac{d i_k(t)}{dt}=\lambda \left[1-i_k(t)\right] k {\Theta}_k(t),
  \label{SIk}
\end{equation}
where the creation term is proportional to the spreading rate $\lambda$, the
degree $k$, the probability $1-i_k$ that a vertex with degree $k$ is
not infected, and the density $\Theta_k$ of infected neighbours
of vertices of degree $k$. The latter term is thus the
average probability that any given neighbor of a vertex of degree $k$
is infected. 

The simplest situation one can encounter corresponds to a complete
lack of correlations (We will address the case of correlated networks
in section 3.4). A network is said to have no degree correlations when
the probability that an edge departing from a vertex of degree $k$
arrives at a vertex of degree $k'$ is independent of the degree of the
initial vertex $k$.  In this case, the probability that each edge of a
susceptible is pointing to an infected vertex of degree $k'$ is
proportional to the fraction of edges emanated from these vertices.
By considering that at least one of the edges of each infected vertex
is pointing to another infected vertex, from which the infection has
been transmitted, one obtains
\begin{equation}
  \Theta_k(t)=\Theta(t)=\frac{\sum_{k'} (k'-1)P(k') i_{k'}(t)}{\avk},
\label{theta}
\end{equation}
where $\langle k \rangle =\sum_{k'} k'P(k')$ is the proper
normalization factor dictated by the total number of edges. At $t=0$,
if the initial density is not too large, all the neighbors of the
initially infected vertices are susceptible and we have
\begin{equation}
  \Theta(t=0)=\frac{\sum_{k'} k'P(k') i_{k'}(0)}{\avk}\ .
\label{theta0}
\end{equation}

A reaction rate equation for $\Theta(t)$ can be obtained from
Eqs.~(\ref{SIk}) and~(\ref{theta}). In the initial epidemic stages,
neglecting terms of order ${\cal O}(i^2)$, the following set of
equations is obtained
\begin{eqnarray}
\frac{d i_k(t)}{dt}&=&\lambda k \Theta(t),\\
\frac{d \Theta(t)}{dt}&=&\lambda \left( \frac{\fluck}{\avk}-1 \right)\Theta(t).
\label{eq_iktheta}
\end{eqnarray}
These equations can be solved, yielding for the prevalence of nodes of
degree $k$, in the case of an uniform initial condition $i_k(t=0) =i_0$,
\begin{equation}
  i_k(t) =i_0 \left[1+\frac{k(\avk-1)}{\fluck -\avk}(e^{t/\tau}-1)\right],
  \label{eq_ik}
\end{equation}
and for the total average prevalence $i(t) = \sum_k P(k) i_k(t)$
\begin{equation}
 i(t) =i_0 \left[1+\frac{\avk^2-\avk}{\fluck-\avk}(e^{t/\tau}-1)\right],
  \label{eq_avik}
\end{equation}
where 
\begin{equation}
  \tau = \frac{\avk}{\lambda (\fluck-\avk)}.
\label{tauSF}
\end{equation}

The result Eq.~(\ref{tauSF}) for uncorrelated networks readily implies
that the growth time scale of an epidemic outbreak is related to the
graph heterogeneity. Indeed, the ratio
\begin{equation}
\kappa=\frac{\fluck}{\avk}
\end{equation}
is the parameter defining the level of heterogeneity of the network,
since the normalized degree variance can be expressed as $\kappa/\avk-1$ and
therefore high levels of fluctuations correspond to $\kappa\gg 1$. In
homogeneous networks with a Poisson degree distribution, in which $\kappa=
\avk+1$ we recover the result $\tau = (\lambda\avk)^{-1}$, corresponding to
the homogeneous mixing hypothesis. On the other hand, in networks with
very heterogeneous connectivity patterns, $\kappa$ is very large and
the outbreak time-scale $\tau$ is very small, signaling a very fast
diffusion of the infection. In particular, in scale-free networks
characterized by a degree exponent $2<\gamma\leq 3$ we have that
$\kappa\sim\fluck\to\infty$ with the network size $N\to\infty$.
Therefore in uncorrelated scale-free networks we face a virtually
instantaneous rise of the epidemic incidence.

%-----------------------------
\Subsection{Extension to the SIS and SIR models}

It is worth stressing that the above results can be easily extended to
the SIS and the SIR models \citep{anderson92}. In the case of
uncorrelated networks, Eq.~(\ref{SIk}) contains, for both the SIS and
the SIR models, an extra term $-\mu i_k(t)$ defining the rate at which
infected individuals of degree $k$ recover and become again
susceptible or permanently immune and thus removed from the
population, respectively:
\begin{equation}
  \frac{d i_k(t)}{dt}=\lambda k s_k(t) \Theta_k(t)-\mu i_k(t) \ .
  \label{SISk}
\end{equation}
In the SIS model we have, as usual, $s_k(t) = 1 - i_k(t)$. In the SIR
model, on the other hand, the normalization imposes that
$s_k(t)=1-i_k(t)-r_k(t)$, where $r_k(t)$ is the density of removed
individuals of degree $k$.  The inclusion of the decaying term $-\mu
i_k$, does not change the picture obtained in the SI model.  By using
the same approximations, the time scale is found to behave as
\begin{equation}
\tau \sim \frac{\avk}{\lambda\fluck-(\mu+\lambda)\avk}
\label{tau_SIS}
\end{equation}
In the case of diverging fluctuations the time-scale behavior is
therefore still dominated by $\fluck$ and $\tau$ is always positive
and going to $0$ as $\fluck \to \infty$
whatever the spreading rate $\lambda$. This allows to recover the
absence of an epidemic threshold, i.e. the lack of a decreasing
prevalence region in the parameter space. It must be noted, however,
that if fluctuations are not diverging, the outbreak time scale is
slightly different for the SIS and SIR models (equation (\ref{tau_SIS}))
than for the SI model (equation (\ref{tauSF})).

%-----------------------------
\Subsection{Heterogeneous correlated networks}

While we have so far restricted our study to the case of uncorrelated
networks, it is worth noting that many real networks do not fill
this assumption \citep{mendesbook,romuvespibook}. In order to consider 
the presence of non trivial correlations we have to fully take into
account the structure of the conditional correlation function
$P(k'|k)$. The equations we have written for the evolution of $i_k$ in
the SI model can therefore be stated as \citep{marianproc} 
\begin{eqnarray}\nonumber
\frac{d i_k(t)}{dt} &=&\lambda \left[1-i_k(t)\right] k \Theta_k(t) \\
 \Theta_k &=&  \sum_{k'} i_{k'} \frac{k'-1}{k'} P(k'|k).
\end{eqnarray}
Here the $\Theta_k$ function takes into account explicitly the
structure of the conditional probability that an infected vertex with degree 
 $k'$ points to a vertex of degree $k$, with any of the $k'-1$ free edges it
has (not pointing to the original source of its infection).
In the absence of correlations, it is possible to see that 
$P(k'|k) = k' P(k')/ \avk$, recovering the results of section 3.2.
If the network presents correlations, measured by $P(k'|k)$, the
situation is slightly more complex. The rate equation for $i_k(t)$ can
be written in this case, neglecting terms of order ${\cal O}(i^2)$, as
\begin{eqnarray}
  \frac{d i_k(t)}{d t} &=& \sum_{k'} \lambda k \frac{k'-1}{k'} P(k'|k)
  i_{k'}(t) \nonumber \\
  &\equiv&  \sum_{k'} C_{k, k'} i_{k'}(t),
\end{eqnarray}
which is a linear systems of differential equations given by the
matrix $\mathbf{C}= \{ C_{k, k'}\}$ of elements
\begin{equation}
   C_{k, k'} = \lambda k \frac{k'-1}{k'} P(k'|k).
\end{equation}
Elementary considerations from mathematical analysis tell us that the
behavior of $i_k(t)$ will be given by a linear combination of
exponential functions of the form $\exp(\Lambda_i t)$, where
$\Lambda_i$ are the eigenvalues of the matrix $\mathbf{C}$. Therefore,
the dominant behavior of the averaged prevalence will be
\begin{equation}
  i(t) \sim e^{\Lambda_mt},
\end{equation}
where $\Lambda_m$ is the largest eigenvalue of the matrix $\mathbf{C}$. In
the case of an uncorrelated network, we have that $C^{\mathrm{nc}}_{k,
  k'} = \lambda k (k'-1) P(k') / \avk$, which has a unique eigenvalue
satisfying
\begin{equation}
\sum_{k'}C_{k,k'}\Psi_{k'}=\Lambda^{\mathrm{nc}}_{m}\Psi_k,
\end{equation}
where $\Lambda^{\mathrm{nc}}_{m} = \lambda (\fluck / \avk -1)$, and where the
corresponding eigenvector is $\Psi_k= k$, thus recovering the previous
result Eq.~(\ref{tauSF}) for this kind of networks.

In the case of correlated networks, it has been shown using the
Frobenius theorem~\citep{frobenius} that the largest eigenvalue is
bounded from below~\citep{boguna:corre}
\begin{equation}
\Lambda_m^2\geq\min_k\sum_{k'}\sum_l(k'-1)(l-1)P(l|k)P(k'|l) \ .
\end{equation}
This equation is very interesting since it can be rewritten as
\begin{equation}
\Lambda_m^2\geq\min_k\sum_{l} (l-1)P(l|k) (k_{nn}(l) - 1) \ .
\label{eq:bound}
\end{equation}
It has been shown \citep{boguna:corre} that, for scale-free network
with $2\leq\gamma\leq 3$, $k_{nn}(l)$ diverges for infinite size
systems ($N\to\infty$). This ensures that also $\Lambda_m$
diverges. Two particular cases have however to be treated separately:
it may happen that, for some $k_0$, $P(l|k_0)=0$; then the previous
limit for $\Lambda_m^2$ gives no information but it is then possible
to show with slightly more involved calculations that $\Lambda_m$
still diverges for $N\to\infty$\citep{marianproc}. Another problem
arises if $k_{nn}(l)$ diverges only for $l=1$; this happens however
only in particular networks where the singularity is accumulated in a
pathological way onto vertices with a single edge. Explicit examples
of this situation are provided in
\citep{vazmo}.

The previous result has the important consequence that, 
even in the presence of correlations, the time scale $\tau\sim
1/\Lambda_M$ 
tends to zero in the thermodynamic limit for any scale-free network with
$2<\gamma\leq 3$. It also underlines the relevance of the
quantity $k_{nn}$, which gives a lower-bound for $\Lambda_m$ in finite
networks. In the next sections we will analyze numerically both the SI
and SIS models in order to provide a full account of the dynamical 
properties that takes into account the network's complexity as well as
finite size effects in the population.

\Section{Numerical simulations}

In order to test  the analytical predictions made in the previous
sections, we have performed extensive numerical simulations of the SI
and the SIS model in two different paradigmatic examples of complex
network models with  homogeneous and heterogeneous properties. 
The choice of the network models was dictated by the request of 
generating an uncorrelated network, to fully exploit the analytical 
predictions. Simulations use an agent-based modeling strategy in 
which at each time step the SI dynamics is applied to each vertex 
by considering the actual state of the vertex and its neighbors.  
It is then possible to measure the evolution of the number of 
infected individuals and other
quantities. In addition, given the stochastic nature of the model,
different initial conditions and networks realizations can be used to
obtain averaged quantities.  In our simulations, we use $N=10^4$ and
$\avk$ ranging from $4$ to $20$. We typically average our results for
a few hundred networks while for each network, we average over a few
hundreds different initial conditions.

\Subsection{Homogeneous networks}

The example of homogeneous complex network we have chosen is the
random graph model proposed by Erd{\"o}s and Renyi
\citep{erdos59,bollobas}. The network is constructed from a set of $N$
different vertices, in which each one of the $N(N-1)/2$ possible edges
is present with probability $p$ (the connection probability), and
absent with probability $1-p$. This procedure results in a random
network with average degree $\avk = pN$ and a Poisson degree
distribution in the limit of large $N$ and constant $\avk$,
\begin{equation}
  P(k) = e^{-\avk} \frac{\avk^k}{k!}.
\end{equation}
For $\avk >1$, it can be proved that the network exhibits a
\textit{giant component}, that is, a set of connected vertices whose
size is proportional to $N$. In an actual realization, however, it is
possible to generate networks in which a fraction of the vertices
belong to disconnected clusters. Therefore, we make the computer
simulations of the spreading models only on the giant component.

In the case of the random graph, we checked the validity of
equation (\ref{i_SI_H}), and that the time scale
is given by $1/\lambda\avk$, as can be seen in Fig.~(\ref{fig:tau_ER}).

\begin{figurehere}
\vspace*{1cm}
\centerline{
\epsfysize=0.5\columnwidth{\epsfbox{fig_SI_ER.eps}}
}
\vspace*{0.1cm}
\caption{ \small 
Main frame: the symbols correspond to simulations of the SI model with
$\lambda=10^{-4}$ on ER networks with $N=10^4$, $\avk=20, 40$; the
lines are fits of the form of Eq. (\ref{i_SI_H}). Inset: measured time
scale $\tau$, as obtained from fitting, versus the theoretical
prediction for different values of $\avk$ and $\lambda$.}
\vspace*{1cm}
\label{fig:tau_ER}
\end{figurehere}
Needless to say, in the case of a homogeneous network, the hypothesis $k \simeq
\langle k \rangle$ captures the correct dynamical behavior of the spreading.
This is a standard result and we report the numerical simulations just
as a reference for comparison with the following numerical
experiments on heterogeneous networks.

%%%%%%%%%%%%%%%%%%%%%%%%%%%%%%%%%%%%%%%%%%%%%%%%%%%%%%%%%%%%
\Subsection{Heterogeneous networks}

As a typical example of heterogeneous network, we have chosen the
networks generated with the Barab{\'a}si-Albert (BA) algorithm
\citep{barab99}. In this algorithm, the network starts from a small
core of $m_0$ connected vertices. At each time step a new vertex is
added, with $m$ edges ($m<m_0$) connected to the old vertices in the
network. The new edges are connected to an old vertex $i$ with a
probability proportional to its degree $k_i$ (\textit{preferential
attachment}). The networks generated by this algorithm have a minimum
degree $m$, an average degree $\avk=2m$, a scale-free degree
distribution $P(k)\sim k^{-3}$ and vanishingly small correlations
\citep{romuvespibook}. In our simulations we use different network
sizes $N$ and minimum degree values $m$ in order to change the level
of heterogeneity, that in this case is given by $\kappa\sim m \ln N$.

In the simulation we keep track of the average density of infected 
individuals versus time for a network generated with the preferential 
attachment rule. In Fig.~\ref{fig:fit} we show a typical result and 
the fitting procedure which allows us to measure the value of $\tau$.

%%%%% figure fit
\begin{figurehere}
\vspace*{0.1cm}
\centerline{
\epsfysize=0.5\columnwidth{\epsfbox{fig1a.eps}}
}
\vspace*{0.2cm}
\caption{ \small Average density of infected individuals versus time in a
BA network of N=$10^4$ with $m=2$.  The inset shows the exponential
fit obtained in the early times (lines) and the numerical curves
$i(t)$ for networks with $m=4,\ 8,\ 12,\ 20$ (from bottom to top).
}
%\vspace*{0.2cm}
\label{fig:fit}
\end{figurehere}

In Fig.~\ref{fig:tau} we report the early time behavior of outbreaks
in networks with different heterogeneity levels and the behavior of
the measured $\tau$ with respect to $\avk/\lambda(\fluck-\avk)$. The
numerical results recover the analytical prediction with great
accuracy.  Indeed, the BA network is a good example of uncorrelated
heterogeneous network in which the approximations used in the
calculations are satisfied. In networks with correlations we expect to
find different quantitative results but a qualitatively similar
framework as it happens in the case of the epidemic threshold
evaluation~\citep{marianproc}.

\begin{figurehere}
\vspace*{1cm}
\centerline{
\epsfysize=0.5\columnwidth{\epsfbox{fig1b.eps}}
}
\vspace*{0.1cm}
\caption{ \small Measured time scale $\tau$ in BA networks as obtained from
exponential fitting versus the theoretical prediction for different
values of $m$ and $N$ corresponding to different levels of
heterogeneity.  }
%\vspace*{1cm}
\label{fig:tau}
\end{figurehere}

%%%%%%
\Subsection{The SIS dynamics}

As claimed in previous sections, we have tested that at short times
and for large heterogeneous networks ($\kappa\gg 1$), the SI model is
a good approximation to the more general SIS model described by
Eq.~(\ref{SISk}). In Fig.~\ref{fig:SIS} we plot $i(t)$ versus $t$ for
a simulation of the SI and the SIS model with $\lambda=10^{-3}$, on a
BA network with $m=8$ and $N=10^4$, corresponding to $\avk = 16$ and
$\fluck \approx 632$. We took different values for the parameter
$\mu$: $\mu=0$ (which corresponds to the SI model), $\mu=10^{-4},
10^{-3}, 10^{-2}$; Eq.~(\ref{tau_SIS}) yields the following values for
$\tau$: $\tau(\mu=0) \approx 26$, $\tau(\mu=0.0001) \approx 26.1$,
$\tau(\mu=0.001) \approx 26.7$, $\tau(\mu=0.01) \approx 35$.
Figure~\ref{fig:SIS} clearly shows that indeed for $\mu \avk \ll
\lambda
\fluck $ and short times the curves are indistinguishable so that the
SIS model is well described by the SI approximation in this regime. As
$\mu$ gets larger the behavior changes since the time scale of the
recovery is no longer much larger than the other time scales. It is
possible to show that that the same results apply also in the case of
the SIR model.

\begin{figurehere}
\vspace*{1cm}
\begin{center}
\epsfig{file=fig_sis_ba_m8.eps,width=8cm}
\end{center}
\caption{\small
SIS model on a BA network with $N=10^4$, $m=8$; $\lambda=10^{-3}$, and
$\mu=0, 10^{-4}, 10^{-3}, 10^{-2}$ from top to bottom. The inset
focuses on the short time dynamics: the curve is superimposed on data
for the SI model up to more than $10\tau(\mu=0)$ for $\mu=10^{-4}$ and
up to a few times $\tau(\mu=0)$ for $\mu= 10^{-3}$.}
\vspace*{1cm}
\label{fig:SIS} 
\end{figurehere}

%%%%%%%%%%%%
\Section{The infection time pattern: the cascade effect}

The previous results show that the heterogeneity of scale-free
connectivity patterns favors epidemic spreading not only by
suppressing the epidemic threshold, but also by accelerating the
epidemic propagation in the population.  The velocity of the spreading
leaves us with very short response times in the deployment of control
measures and a detailed knowledge of the way epidemics propagate
through the network could be very valuable in the definition of
adaptive strategies. Indeed, the epidemic diffusion is far from
homogeneous.  The simple formal integration of Eq.~(\ref{SIk})
written for $s_k=1-i_k$ yields
\begin{equation} 
s_k(t)=s_{k}^0e^{-\lambda k\Phi(t)} \ ,
\end{equation}
where $\Phi(t)=\int_0^t d t' \Theta(t')$.  This result is valid for
any value of the degree $k$ and the function $\Phi(t)$ is positive and
monotonously increasing. This last fact implies that $s_k$ is
decreasing monotonously towards zero when time grows. Thus, if one has
two values $k>k'$ and whatever the initial conditions $s_{k}^0$ and
$s_{k'}^0$, there is a time $t_\times$ after which $s_k(t)<s_{k'}(t)$.
This `crossing' time is given by
\begin{equation}
\phi(t_\times)=
\frac{1}{\lambda(k-k')}\log\left[\frac{s_{k}^0}{s_{k'}^0}\right] \ .
\end{equation}
(If $s_{k}^0<s_{k'}^0$, there is no crossing and for all times
$s_k(t)<s_{k'}(t)$). This result indicates that after an initial
regime which depends on the initial conditions, the disease spreads
always from large connectivity to smaller connectivities.

We can go beyond the analytical result that is necessarily grounded on
average quantities by relying on numerical simulations. Indeed, a 
more precise characterization of the epidemic diffusion through 
the network can be achieved by studying some convenient quantities 
that highlight the invasion pattern of the infection in numerical 
spreading experiments in BA networks. First, we measure the
average degree of the newly infected nodes at time $t$, defined as
\begin{equation}
\overline{k}_{\mathrm{inf}}(t)=
\frac{\sum_k k[I_k(t)-I_k(t-1)]}{I(t)-I(t-1)} \ .
\label{knew}
\end{equation}
In Fig.~\ref{fig:kinf} we plot this quantity for BA networks as a
function of the rescaled time $t/\tau$. The curves show an initial
plateau that can be easily understood by considering that at very low
density of infected individuals, each vertex will infect a
fraction of its neighbors without correlations with the spreading from
other vertices. In this case each edge points to a vertex with degree
$k$ with probability $kP(k)/\avk$ and the average degree of newly
infected vertices is given by
\begin{equation}
\overline{k}_{\rm inf}(t) =\fluck/\avk \ .
\end{equation}
After this initial regime, $\overline{k}_{\rm inf}(t)$ decreases
smoothly when time increases. The dynamical spreading process is
therefore clear: after the hubs are very quickly infected the spread
is going always towards smaller values of $k$. This is confirmed by
the large time regime that settles in a plateau
\begin{equation}
\overline{k}_{\rm inf}(t) = m \ ,
\end{equation}
which means that the vertices with the lowest degree are typically the
last to be infected.

%%%%% figure kinf
\begin{figurehere}
\vspace*{1cm}
\centerline{
\epsfysize=0.5\columnwidth{\epsfbox{fig2a.eps}}
}
\vspace*{0.1cm}
\caption{ \small Time behavior of the average degree of the newly
infected nodes for SI outbreaks in BA networks (here of size $N=10^4$).
Time is rescaled by $\tau$. Reference lines are drawn at the
asymptotic values $\fluck/\avk$ for $t\ll\tau$ and $m$ for
$t\gg\tau$. The two curves are for $m=4$ (bottom) and $m=14$ (top).}
\label{fig:kinf}
\vspace*{1cm}
\end{figurehere}

Further information on the infection propagation is provided by the
inverse participation ratio $Y_2(t)$
\citep{Derrida:1987,barthelemy02}. We first define the weight of
infected individuals in each degree class $k$ by
$w_k(t)=I_k(t)/I(t)$. The quantity $Y_2$ is then defined as
\begin{equation}
  Y_2(t)=\sum_k w_k^2(t).
\label{y2}
\end{equation}
If $Y_2\sim 1/k_{max}$ ($k_{max}$ is the maximal connectivity),
infected vertices are homogeneously distributed among all degree
classes. In contrast, if $Y_2$ is not small (of order $1/n$ with $n$
of order unity) then the infection is localized on some specific
degree classes that dominate the sum of Eq.~(\ref{y2}). In
Fig.~\ref{fig:Y2} we report the behavior of $Y_2$ versus time for BA
networks with different minimum degree.  The function $Y_2$ has a
maximum at the early time stage, indicating that the infection is
localized on the large $k$ classes, as we infer from the plot of
$\overline{k}_{\rm inf}(t)$, see Fig.~\ref{fig:kinf}.  First $Y_2$
decreases, with the infection progressively invading lower degree
classes, and providing a more homogeneous diffusion of infected
vertices in the various $k$ classes.  Finally, the last stage of the
process corresponds to the capillary invasion of the lowest degree
classes which have a larger number of vertices and thus provide a
larger weight.  In the very large time limit, when the whole network
is infected, $Y_2(t=\infty)=\sum_kP(k)^2$.  Noticeably, curves for
different levels of heterogeneity have the same time profile in the
rescaled variable $t/\tau$.  This implies that, despite the various
approximations used in the calculations, the whole spreading process
is dominated by the time-scale defined in the early exponential regime
of the outbreak.

The presented results provide a clear picture of the infection
propagation in heterogeneous networks. First the infection takes
control of the large degree vertices in the network.  Then it rapidly
invades the network via a cascade through progressively smaller degree
classes. The dynamical structure of the
spreading is therefore characterized by a hierarchical cascade from
hubs to intermediate $k$ and finally to small $k$ classes. 

%%%%% figure Y2
\begin{figurehere}
\vspace*{1cm}
\centerline{
\epsfysize=0.5\columnwidth{\epsfbox{fig2b.eps}}
}
\vspace*{0.1cm}
\caption{ \small Inverse participation ratio $Y_2$
  versus time for BA network of size $N=10^4$ with minimum degree
  $m=4,6,8,10,12,14$ and 20, from top to bottom. Time is rescaled with $\tau$.
  The reference line indicates the minimum of $Y_2$ around $t/\tau\simeq 6.5$.
  }
\vspace*{1cm}
\label{fig:Y2}
\end{figurehere}
%%%%%

%%%%%%%%%%%

\Section{The effect of the initial seed of the infection}

In very heterogeneous networks it is reasonable to forecast that
epidemic outbreaks starting from individuals possessing very different
connectivity properties may undergo a rather different time evolution.
In order to investigate in more details the effect of initial
conditions, one can write the general solution of
Eqs.~(\ref{eq_iktheta}) with $i_k(t=0)=f(k)$ where $f$ is some given
function of the degree
\begin{equation}
i_k(t)=f(k)+ k \Theta_0\frac{\avk}{\fluck-\avk}\left[e^{t/\tau}-1\right] \ ,
\label{eq:2}
\end{equation}
where $\Theta_0=\Theta(t=0)=\langle k f(k)\rangle /\avk$.
Comparing this last expression with Eq.~(\ref{eq_ik}), we conclude
that, apart from the effect of the initial density of infected
individuals, the only effect in the dynamics is the presence of the
prefactor $\Theta_0$, which implies that the spreading is faster
for larger $\Theta_0$. In the particular case in which the initial
infection is located on $N_0$ vertices of given degree $k_0$, we
obtain $i_{k_0}(t=0)=N_0/(NP(k_0))$ and thus
$\Theta_0=k_0 N_0/(N\avk)$. Analogously it is easy to see that, for
$N_0$ initially contaminated sites, all with connectivity larger than
a certain $k_M$, $\Theta_0$ is proportional both to $N_0/N$ and to
$k_M$.

Eq. (\ref{eq:2}) clearly highlights the role of the large connectivity
sites: on the one hand, for a given initial condition, the infection
of classes with larger $k$ leads to faster rise of the epidemic; on
the other hand, the larger the degree of the initially infected sites,
the faster the propagation of the epidemics. Note however that the
time-scale $\tau$ itself is not affected by the initial conditions,
which appear through the prefactors only.  Using Eq.~(\ref{eq:2}) we
can highlights these features since the right hand side of the
expression
\begin{equation}
  \frac{i_k(t) -f(k)}{k \Theta_0} =
  \frac{\avk}{\fluck-\avk}\left[e^{t/\tau}-1\right] 
\end{equation}
has to  be independent on both the initial conditions (which enter
$i_k$ only through $f(k)$ and $\Theta_0$) and the degree
$k$. We have verified this in Fig.~\ref{fig:initials}, by means of
numerical simulations of the SI model on BA networks with
specific initial conditions: if the spreading starts from $N_0$ infected 
sites of connectivity $k_0$, $\Theta_0$ is proportional
to $N_0$ and $k_0$, so that we show that the quantity
\begin{equation}
\frac{i_k(t) -f(k)}{k N_0 k_0} 
\label{eq:IC1}
\end{equation}
is independent of $k$ and $k_0$, and is also equal to
\begin{equation}
\frac{i(t) - i_0}{\avk N_0 k_0}  \ .
\label{eq:IC2}
\end{equation}
At large times the curves separate since the expression (\ref{eq_ik})
is no longer valid. The data collapse of Fig.~\ref{fig:initials} once
again shows the validity of the analytical approach developed in
section 3. The time scale $\tau$ for the exponentially fast outbreak
of the epidemics is relevant for any kind of initial conditions and
for all nodes, independently from their connectivity. The prefactors
however show that the spreading is faster if the infected seeds have a
large connectivity and, among the connectivity classes, hubs are
typically infected faster.

\begin{figurehere}
\vspace*{1cm}
\centerline{
\epsfysize=0.5\columnwidth{\epsfbox{fig_SI_BA_IC.eps}}
}
\caption{Collapse plot of quantities (\ref{eq:IC1}) and (\ref{eq:IC2}) vs. time
for the SI model on a BA network, with
initial conditions given by $N_0$ infected 
sites having a given connectivity $k_0$. Here $N_0=10$, $N=10^4$.
The data are averaged over $200$ realizations of the network and
$50$ runs on each network sample.
Lines correspond to (\ref{eq:IC1}) for $k_0=10, 20, 30, 50$; symbols
correspond to (\ref{eq:IC2}) for the same initial conditions and various
values of $k$ between $10$ and $90$ (The maximal connectivity is of order 
$\sqrt{N}=100$).}
\label{fig:initials}
\end{figurehere}

\Section{Fluctuations: 
the relevance of $\tau$ for  single case studies}

The previous results are valid for the average density of infected
nodes and it is legitimate to ask about the relevance of these results
to the case of a single network sample. Indeed, in the real-world
there is not such a thing as averages over different network
realizations and we have to check the robustness of our results on a
single network sample. It is then natural to wonder if large
statistical fluctuations in the network structure may reverberate on
the difference between single case studies and asymptotic average
results. In particular, we can compare the previously defined
time scale $\tau$ (Eq. \ref{tauSF}), which corresponds to the
exponential growth of the prevalence averaged over the networks
statistical properties and initial conditions, with an operative
measure of time-scale obtained in a single network realization case
study.  For a given network, and starting from a given initial density
of infected vertices $i_0$, we define for each run of the spreading process the
typical realization time-scale as the time $\tau'$ at which the
density of infected nodes $i_{run}(t)$ in the specific run is equal to
the average quantity $i(t)$ at time $\tau$ provided by
Eq. (\ref{eq_avik}).  This defines $\tau'$ as the effective time at
which $i_{run}(t=\tau')= i(t=\tau)$ (see Figure \ref{fig:tauprime}).
The deviation from the average behavior in each run can be readily
accounted by the variable $u=\tau'/\tau$ which is a simple measure of
the relevance of $\tau$ for each run. In Fig.~\ref{fig:CI}(a), we plot
the probability distribution of $u$ computed for a given realization
of the network.  This figure shows that $P(u)$ is not broadly
distributed.  In other words it is very unlikely to find large
deviations of $u$ from the unity, at which the distribution is
strongly peaked. This evidence indicates that even if fluctuations are
possible, the time-scale defined by $\tau$ is a meaningful
characterization of the spreading process.

\begin{figurehere}
\vspace*{1cm}
\centerline{
\epsfysize=0.5\columnwidth{\epsfbox{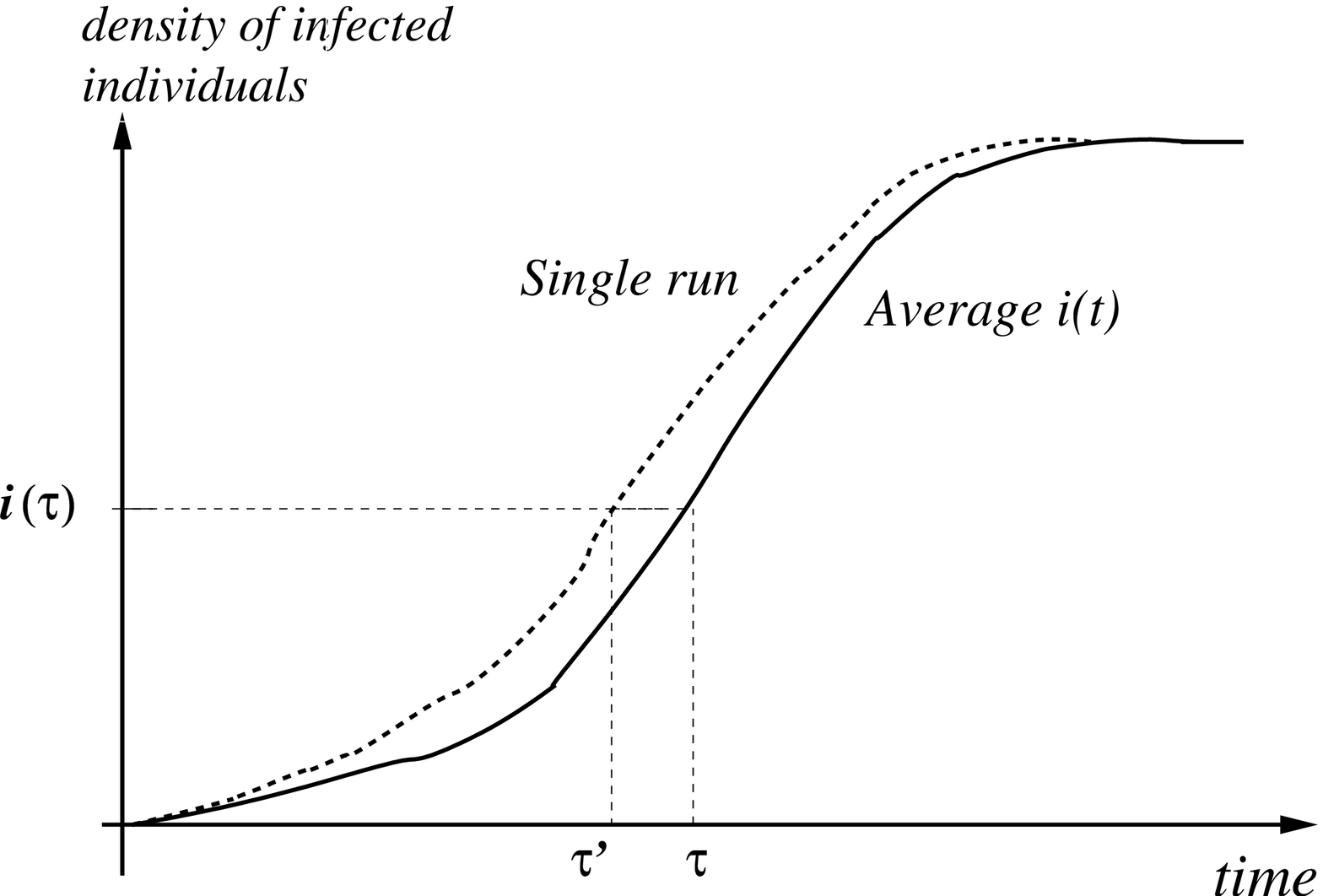}}
}
\vspace*{0.1cm}
\caption{ \small Schematic definition of $\tau'$ for a single run. Given
the average density $i(t)$, $\tau'$ is defined as the time
in which the density of a given run reaches $i(\tau)$.
   } \vspace*{1cm}
\label{fig:tauprime}
\end{figurehere}

%%%%%%

A more detailed analysis highlights once more the key role of initial
conditions in the epidemics spread.  In Fig.~\ref{fig:CI}(b), we used
the same network as in Fig.~\ref{fig:CI}(a), but we made the following
distinction according to the connectivity of the initially infected
nodes. We used initial conditions (i) with small connectivity $k<10$
and (ii) with large connectivity $k>50$ (the maximal connectivity of
the network is $100$ in this case). Figure \ref{fig:CI}(b) shows that
the condition (ii) is the main contribution to the small $u<1$ values
while it is the condition (i) which contributes mostly to the larger
values $u>1$. These results can be easily explained as follows: We
know from previous sections that the first steps of the epidemics
process are concerning the hubs' infection. In the case (i), the path
from the infected seeds to the hubs can be long and can thus induce a
slower growth in $i(t)$.  In the opposite case (ii), instead, the
seeds are mainly located on the hubs since the beginning and the
initial outbreak is generally faster that the average.  This clear
difference tells us that statistical deviations from the average time
scale $\tau$ can be traced back to the initial conditions of the
epidemics.  In this perspective, the specific network realizations
appear to have a rather small influence on statistical
fluctuations. This last feature is due to the intrinsic large amount
of connectivity fluctuations that each network realization possesses in
the case of a very heterogeneous topology. Additional fluctuations, such
as those stemming from the specific realization and finite size, are
thus a higher order correction to the intrinsic statistical
fluctuations which are accounted for correctly in the theoretical
description.

%%%%% figure CI
\begin{figurehere}
\vspace*{1cm}
\centerline{
\epsfysize=0.4\columnwidth{\epsfbox{fig_histtau.eps}}
\hspace*{0.1cm}
\epsfysize=0.4\columnwidth{\epsfbox{fig_histtau.vs.k.eps}}}
\vspace*{0.1cm}
\caption{ \small Probability distribution of $u=\tau'/\tau$. (a) For a
given network with random initial conditions (the BA network is
obtained for $N=10^4$, $m=2$, and $1,000$ realizations). (b)
Probability distribution of $u$ for different initial condition: 
in full lines
the nodes initially infected have a connectivity $k<10$ and in
dashed lines $k>50$ (here $k_{max}=100$).  }
\vspace*{1cm}
\label{fig:CI}
\end{figurehere}

More generally, these results show that initial conditions induce
fluctuations in the outbreak time which are never `large' in the sense
that $P(u)$ decays rapidly, which implies that the time scale $\tau$
is relevant even for a single realization of the network. This result
is important since in real world situations, the disease spreads on one
single network and it has to be checked whether results obtained on
average are relevant or not.

%%%%%%%%%%%%%%%%%%%%%%%%% summary and conclusions
\Section{Conclusions}

In this paper we have provided a general picture of the effect of 
complex connectivity patterns in the spreading dynamics of epidemic
phenomena. We have shown how the large connectivity fluctuations
present in a large class of population networks lead to a novel
epidemiological framework in which both the epidemic threshold and the 
growth time-scale of the outbreak do not have an intrinsic value. 
Indeed, the virtual lack of epidemic threshold and instantaneous rising
of the prevalence in infinitely large networks corresponds to 
non-intrinsic quantities depending on the specific system size of the
actual population. In addition, the dynamical structure of the 
spreading process is characterized by a hierarchical cascade from hubs to 
intermediate $k$ and finally to small $k$ vertices. That is, the 
infection first takes place on the subset of individuals with the 
largest number of contacts, and then progressively invades 
individuals with decreasing number of contacts.  
The emerging picture might be of practical importance in the 
implementation  and assessment of dynamic control strategies. 
In particular, an efficient way to stop epidemics
could rely on a dynamical deployment of containment measures that
focuses on progressively changing classes of the population.
More specifically, our results confirm the importance of control strategies 
targeting the hubs of the population but also highlight
the fact that global surveillance is a major key aspect of epidemics
control and that immunization strategies have to evolve with time
during the different phases of the spread.

%%%%%%%%%%%%%%%%%%%%%%%%%%%%%%%%%%%%%%%%%%%%%%%%%%%%%%

\vspace*{0.5cm}

{\small
  
  A.B, A. V. and R. P.-S. are partially funded by the European
  Commission - Fet Open project COSIN IST-2001-33555. R.P.-S.
  acknowledges financial support from the Ministerio de Ciencia y
  Tecnolog{\'\i}a (Spain), and from the Departament d'Universitats,
  Recerca i Societat de la Informaci{\'o}, Generalitat de Catalunya
  (Spain).}

\vspace*{0.5cm}

{\small

\renewcommand{\baselinestretch}{0.9}

}

\end{multicols}

\end{document}